# Intrinsic Curie temperature bistability in ferromagnetic semiconductor resonant tunneling diodes

Swaroop Ganguly,[1,*] A.H.MacDonald,[2] L.F.Register,[1] and S.Banerjee[1]

[1]Microelectronics Research Center, University of Texas at Austin, TX 78758, USA.

[2]Department of Physics, University of Texas at Austin, TX 78712, USA.

We predict bistability in the Curie temperature-voltage characteristic of double barrier resonant-tunneling structures with dilute ferromagnetic semiconductor quantum wells. Our conclusions are based on simulations of electrostatics and ballistic quantum transport combined with a mean-field theory description of ferromagnetism in dilute magnetic semiconductors.

It has been known for some time now that resonant tunneling diodes (RTD) exhibit hysteresis in their current voltage (I-V) characteristics in the negative-differential-resistance (NDR) region. This effect was first observed by Goldman, Tsui and Cunningham[1] and explained on the basis of charge accumulation in the well region.[2,3,4] Recently, a combination of analytical theory and numerical simulations were used to predict that a RTD with a dilute magnetic semiconductor (DMS) well will undergo a switching of its ferromagnetic critical temperature $T_C$ with applied bias voltage V.[5] In this communication, we show that this structure should also exhibit a hysteresis in its $T_C$-V characteristic. To illustrate this effect we present the results of self-consistent simulations of ballistic quantum transport in the effective mass approximation with Coulomb interactions accounted for in the Hartree approximation, combined with linearized mean field equations for DMS ferromagnetism.

We simulate a large-area device consisting of a highly p-doped GaAs emitter and collector, $Al_{0.8}Ga_{0.2}As$ (band-offset, effective mass and permittivity parameters from Davies[6]) barriers 10Å in thickness, and a 10Å thick (Ga,Mn)As well. The acceptor concentration in the emitter and collector is assumed to be $1x10^{20}cm^{-3}$, while that in the barriers is assumed to be $2x10^{19}cm^{-3}$. We assume an acceptor doping concentration of $4x10^{19}cm^{-3}$ in the well, corresponding to 80% compensation.[7]

The numerical simulations follow the treatment in Ref. 5 – the model equations are described below. Quantum transport is described here using the Keldysh formalism.[8] The equilibrium growth direction (or longitudinal, i.e. parallel to transport) Hamiltonian in a single-band tight-binding model is given by:

$$H_{L\downarrow,\uparrow}\left|z_n\right\rangle = \left(2t + \Delta E_V + U \pm \Delta/2\right)\cdot\left|z_n\right\rangle - t\cdot\sum_{\delta}\left|z_{n+\delta}\right\rangle \qquad (1)$$

$\Delta E_V$ is the valence band offset, $U$ is the electrostatic (Hartree) energy, $\Delta$ is the kinetic exchange potential due to the magnetic impurities, $t \equiv \hbar^2/2m_l^* a^2$ is the hopping energy, $n$ is the site index and $\delta = \pm 1$. The mesh spacing $a$ used here is 1Å. The retarded Green functions for spin-down and spin-up carriers are then given by the Dyson equation:

$$G_{\uparrow,\downarrow}(E) = \left[E - H_{L\uparrow,\downarrow} - \Sigma_e - \Sigma_c\right]^{-1} \qquad (2)$$

where $E$ is the carrier energy and $\Sigma_e, \Sigma_c$ are the self-energies due to coupling to the emitter and collector which are evaluated for the simple 1-D problem considered here by assuming outgoing plane wave boundary conditions.[8] The broadening functions for the emitter and collector are defined by:

$$\Gamma_{e,c} = i \cdot \left( \Sigma_{e,c} - \Sigma_{e,c}^{\dagger} \right) \tag{3}$$

The spectral functions due to the emitter and collector then follow from the steady-state Keldysh equation:

$$A_{e,c} = G \cdot \Gamma_{e,c} \cdot G^{\dagger} \tag{4}$$

The spectral functions and the Green functions above are implicitly spin-dependent. We do not, however, take into account any spin-flip scattering in the ballistic limit considered here. We assume constant chemical potentials $\mu_e$ and $\mu_c$ deep inside the emitter and collector, with $V = \mu_e - \mu_c$ being the applied bias. Then, the non-equilibrium density matrices for spin-up and spin-down carriers are obtained from the expression:

$$\rho_{\uparrow,\downarrow} = \int_{-\infty}^{\infty} \frac{dE}{2\pi} \left[ F_0 \left( E - \mu_e \right) \cdot A_{e\uparrow,\downarrow} + F_0 \left( E - \mu_c \right) \cdot A_{c\uparrow,\downarrow} \right] \tag{5}$$

where $F_0 \left( E - \mu \right) = \sum_{\kappa} f_0 \left( E + \varepsilon_{\kappa} - \mu \right) = S \cdot \frac{m_t^* k_B T}{2\pi\hbar^2} \cdot \ln \left[ 1 + \exp \left( \frac{\mu - E}{k_B T} \right) \right]$ is the sum of the Fermi occupation probabilities for any one spin over all 2D transverse (perpendicular to transport) wavevectors $\kappa$. Here, $S$ is the transverse cross sectional area, and $\varepsilon_{\kappa} = \frac{\hbar^2 \kappa^2}{2 m_t^*}$ are the plain wave energy eigenstates in the transverse direction. The spin-up and spin-down carrier densities in the quantum well are then given by:

$$\Omega \cdot p_{\uparrow,\downarrow} \left( z \right) = \left[ \rho_{\uparrow,\downarrow} \left( z, z' \right) \right]_{z'=z} \tag{6}$$

where $\Omega = a$ , is in general, the volume of an individual cell.

The Hartree potential energy $U$ is obtained from the Poisson equation:

$$\frac{d}{dz} \left( \varepsilon(z) \frac{dU}{dz} \right) = q^2 \cdot \left[ p - N_A \right] = q^2 \cdot \left[ p_{\uparrow} + p_{\downarrow} - N_A \right] \tag{7}$$

$p_{\uparrow}$, $p_{\downarrow}$ being the concentrations of spin-up and spin-down holes, and $N_A$ the acceptor ion concentration. The Poisson equation with the Neumann boundary condition, is solved self-consistently with Eqs. (1)-(5) using Newton-Raphson iteration.

The mean polarization of a magnetic ion in the absence of an external field is given by the following expression in the mean-field picture approximation: [9]

$$\langle M\rangle_I = S\cdot B_S\left(J_{pd}\cdot S\cdot\left[p_\uparrow(R_I)-p_\downarrow(R_I)\right]/2k_BT\right) \tag{8}$$

where $B_S$ is the Brillouin function given by:

$$B_S(x)=\frac{2S+1}{2S}\coth\left(\frac{2S+1}{2S}\cdot x\right)-\frac{1}{2S}\coth\left(\frac{1}{2S}\cdot x\right)$$

$$\xrightarrow[x<<1]{}\frac{S+1}{3S}\cdot x, \tag{9}$$

$S$ = 5/2 is the magnetic ion spin, and $J_{pd}$, the antiferromagnetic exchange coupling between the itinerant p-like valence band electrons and the half-filled d-shell electrons, is assigned a typical experimental value, 150meV.nm$^3$.[9] Then the spin-dependent kinetic-exchange potential is obtained, in the continuum limit, from:

$$\Delta(z)=J_{pd}\cdot N_{Mn}(z)\cdot\langle M\rangle(z) \tag{10}$$

$N_{Mn}$ being the total Mn atom concentration.

Linearizing Eq. (8) using Eq. (9), and continuing to non-zero bias voltage, we have:

$$p_\uparrow(z;V)-p_\downarrow(z;V)=\frac{6k_BT}{S(S+1)J_{pd}}\cdot\langle M\rangle(z;V)=\frac{6k_BT}{S(S+1)J_{pd}^2N_{Mn}}\cdot\Delta(z;V). \tag{11}$$

Note that the above linearization is valid in the limit of small magnetization and spin density, and this limit defines the critical temperature. Lee et al.[9] have shown that for a DMS quantum well in the neighborhood of the critical temperature, the magnetization $\langle M\rangle(z)\propto|\varphi(z)|^2$, the probability density in the well. It has also been shown that the simplest qualitative features of quantum transport across the ferromagnetic semiconductor RTD can be correctly reproduced from the approach of Lee et al. by approximating the voltage-dependent RTD wavefunction in the well region $\psi(z;V)$ by quantum well wavefunctions $\varphi(z)$.[5] With this in mind, we follow the following scheme to *estimate* the Curie temperature $T_C$. After a self-consistent density matrix is found for the paramagnetic state, we introduce a small spin-splitting potential $\Delta(z)$. The spatial variation of the potential is assumed to be proportional to the carrier density $p(z)$; that is, $\Delta(z) \propto |\psi(z;V)|^2 \approx c.p(z;V)$. The corresponding spin density is calculated, once again from a self-consistent calculation. Thus, we may write Eq. (11) as follows:

$$L(\Delta;V) \cong p_\uparrow(z;V) - p_\downarrow(z;V) = \frac{6k_B T}{S(S+1)J_{pd}^2 N_{Mn}} \cdot \Delta(z;V) \qquad (12)$$

where the functional $L$ maps a small $\Delta(z)$ to $p_\uparrow(z) - p_\downarrow(z)$ via a solution of the quantum transport and Poisson equations. Finally, the peak spin density – that at the center of the well $\bar{z}$, is used to calculate the voltage-dependent Curie temperature from Eq. (12) as follows:

$$T_C(V) \approx \frac{S(S+1)J_{pd}^2 N_{Mn}}{6k_B} \left(\frac{\delta L}{\delta \Delta}\right)_{z=\bar{z},c\to 0} \qquad (13)$$

We have increased the bias from zero in steps of 30mV. An initial guess is required for the electrostatic potential at equilibrium. At each subsequent bias point, the converged electrostatic potential for the previous point serves as the initial guess. Fig. 1 shows the I-V characteristic of the device obtained from a simulation of the paramagnetic state, that is, for zero spin-splitting. For a high enough bias, the current falls sharply as expected for a RTD – this occurs when the resonance energy level in the well is pulled down below the band-edge on the emitter side. Beyond this point we have decreased the voltage, with the same step-size, back to zero. We see a hysteresis in the I-V as expected.[1,2,3,4] Fig. 2 illustrates the origin of the bistability by considering the two stable solutions for the 280mV bias point. We note that as expected[3] the current carrying solution has the resonant level above the emitter band-edge with the concomitant (net positive) space-charge accumulation, whereas in the situation without current, the bottom of the well is lowered, pulling down the resonance below the emitter band-edge and reducing the charge build-up drastically to a point of substantial depletion (i.e., net negative space-charge). The vicissitude of charge storage in the well is especially large because of the heavy doping in the leads and the well, resulting in a hysteresis region much wider than in Ref. 3. We note that high Mn doping (in the well) is a prerequisite for DMS ferromagnetism.

Finally Fig. 3 shows our results for the $T_C$-V characteristics which exhibit a pronounced bistability. At a given V the critical temperature $T_C$ has been evaluated using equation (13) above. The simplest analytical theory for this device[5] predicts that for increasing bias, the $T_C$ will drop in two steps, first to about half its equilibrium value at $V \approx V_S \cong 2(\mu - \varepsilon)$ when the resonance energy level is pulled down below the chemical potential on the collector side, and then to zero at $V \approx V_C \cong 2\varepsilon$, when the resonance level is pulled down below the band-edge on the emitter side; $\mu$ being the equilibrium chemical potential and $\varepsilon$ the resonance energy level at equilibrium. In this case, $\mu = 0.15eV$ and $\varepsilon = 0.13eV$. According to this picture, the first step should be at $V_S = 40mV$ and is found in fact almost exactly where expected. The second step is expected to occur near a bias of $V_C = 260mV$ but this value is actually found to be approximately half-way between $V_C$ for the forward and reverse paths. This deviation from the simple analytic result is due to the fact that the analytical theory does not take electron-electron interactions into account. Predictably, it therefore gives a value somewhere between the highly positive and highly negative space-charge scenarios.

In conclusion, we have predicted a novel spin hysteresis effect in ferromagnetic semiconductor resonant tunneling diodes, bistability in the $T_C$-V characteristics. We have explored the systematics of this effect by calculating $T_C$ in a mean field approximation. The critical temperature is determined by linearizing response to a small spin-splitting field of the solution to the transport and electrostatics equations. The effect follows from the well-understood 'charge hysteresis' observed in a classical RTD. Although the width of the hysteresis obtained from our simulations reflects the stability region of bistable solutions rather than what might be quantitatively expected from experiment,[3] our calculations nevertheless indicate that $T_C$-V hysteresis might be an observable effect in ferromagnetic well resonant tunneling diodes. The effect could be verified directly by magnetic or optical circular dichroism measurements.

This work was supported in part by the MARCO MSD Focus Center, NSF, and the Texas Advanced Technology Program. A.H.M. was supported by the Welch Foundation, by the Department of Energy under grant DE-FG03-02ER45958, and by the DARPA SpinS program. S.G. would like to thank Prof. Marc Cahay for stimulating discussions.

REFERENCE

[*] Electronic Address: swaroop@ece.utexas.edu

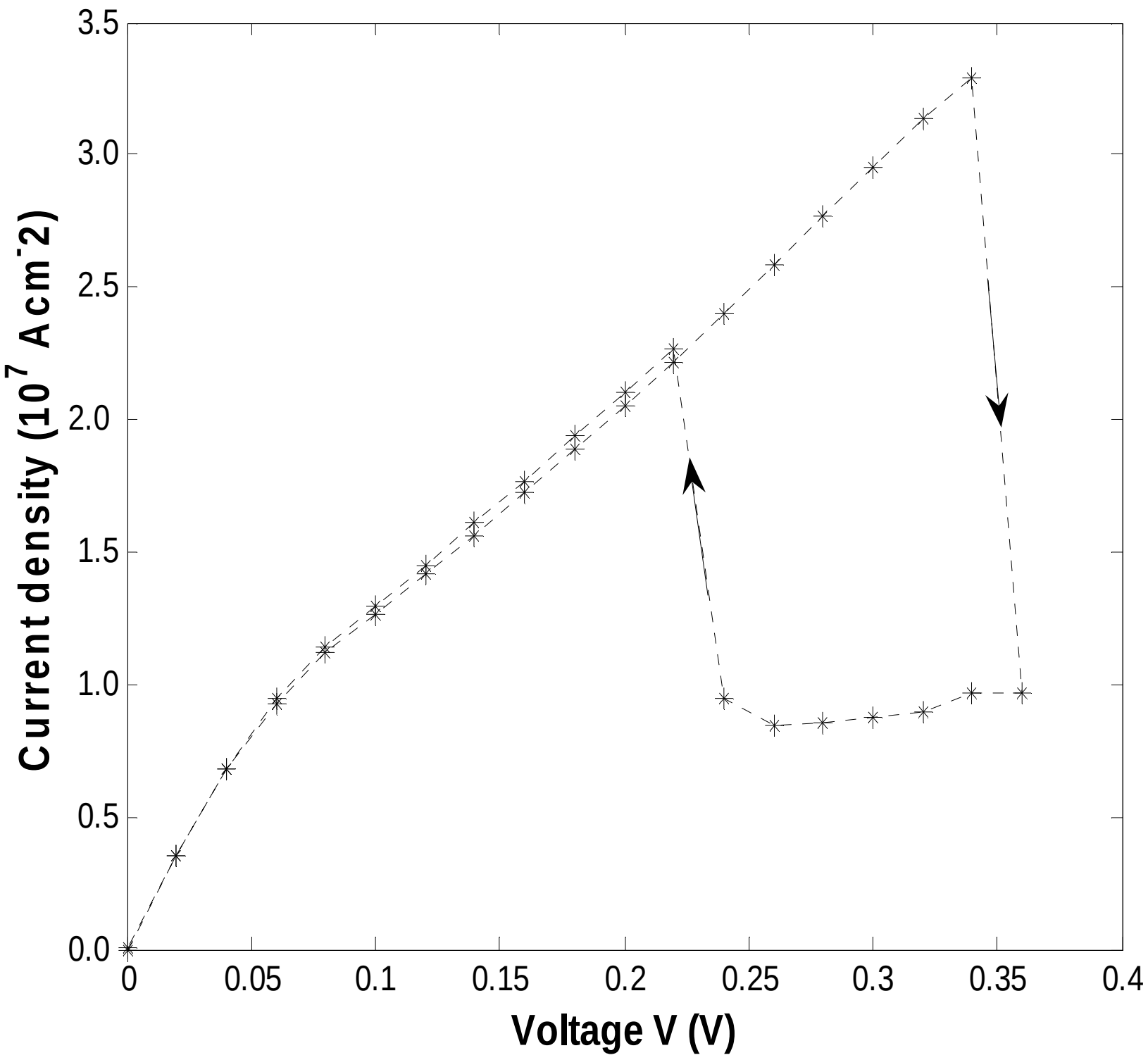

FIG. 1. Hysteretic current voltage (I-V) characteristics of the ferromagnetic well resonant tunneling diode.

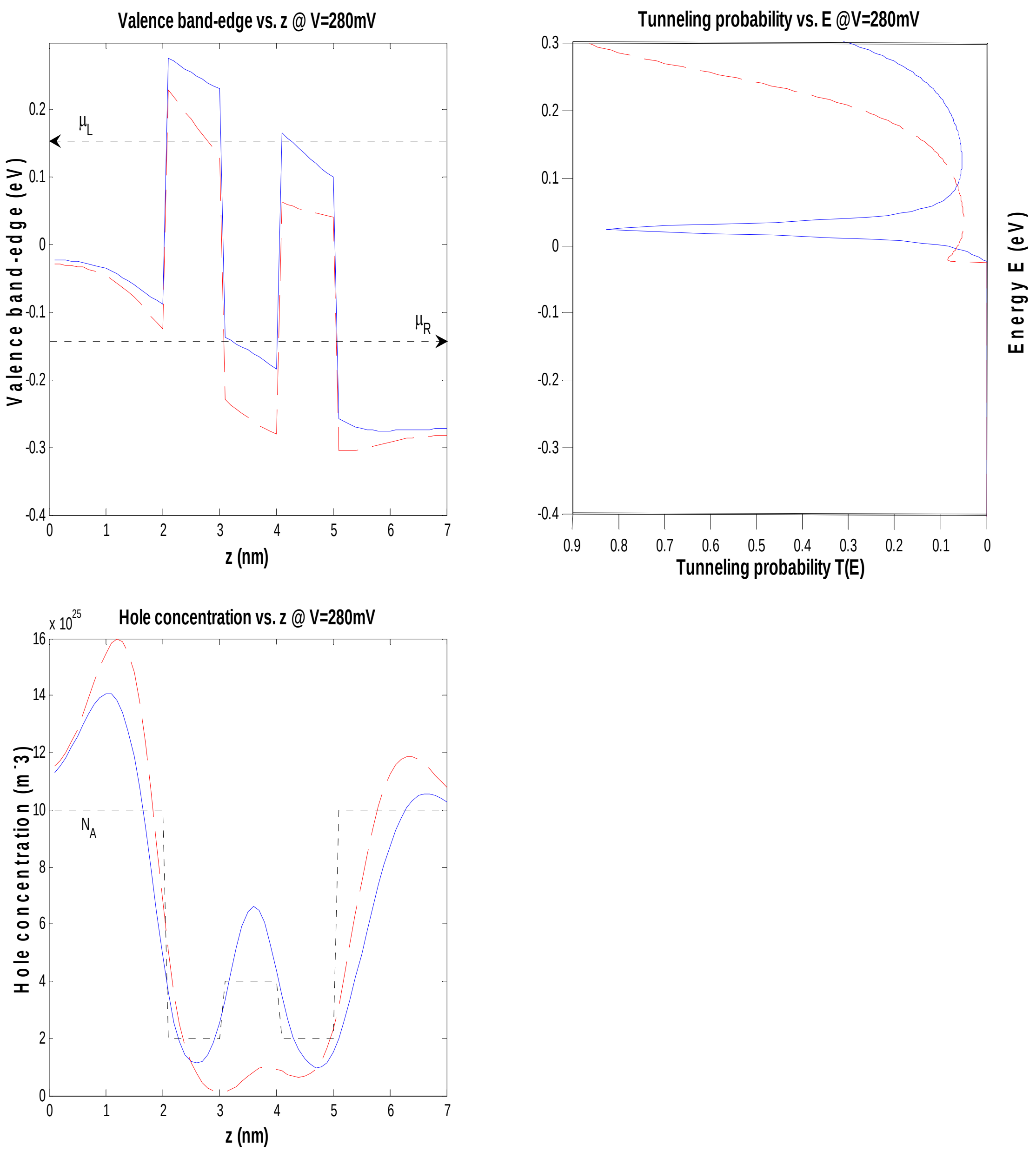

FIG. 2. (Color Online) Paramagnetic state: V=280mV; solid (blue) lines denote the forward path (increasing voltage) and broken (red) lines denote reverse path (decreasing voltage). The former is the current-carrying state with large charge accumulation. In dotted lines: $\mu_L$ , $\mu_R$ are the chemical potentials in the left and right leads, $N_A$ is the acceptor doping profile.

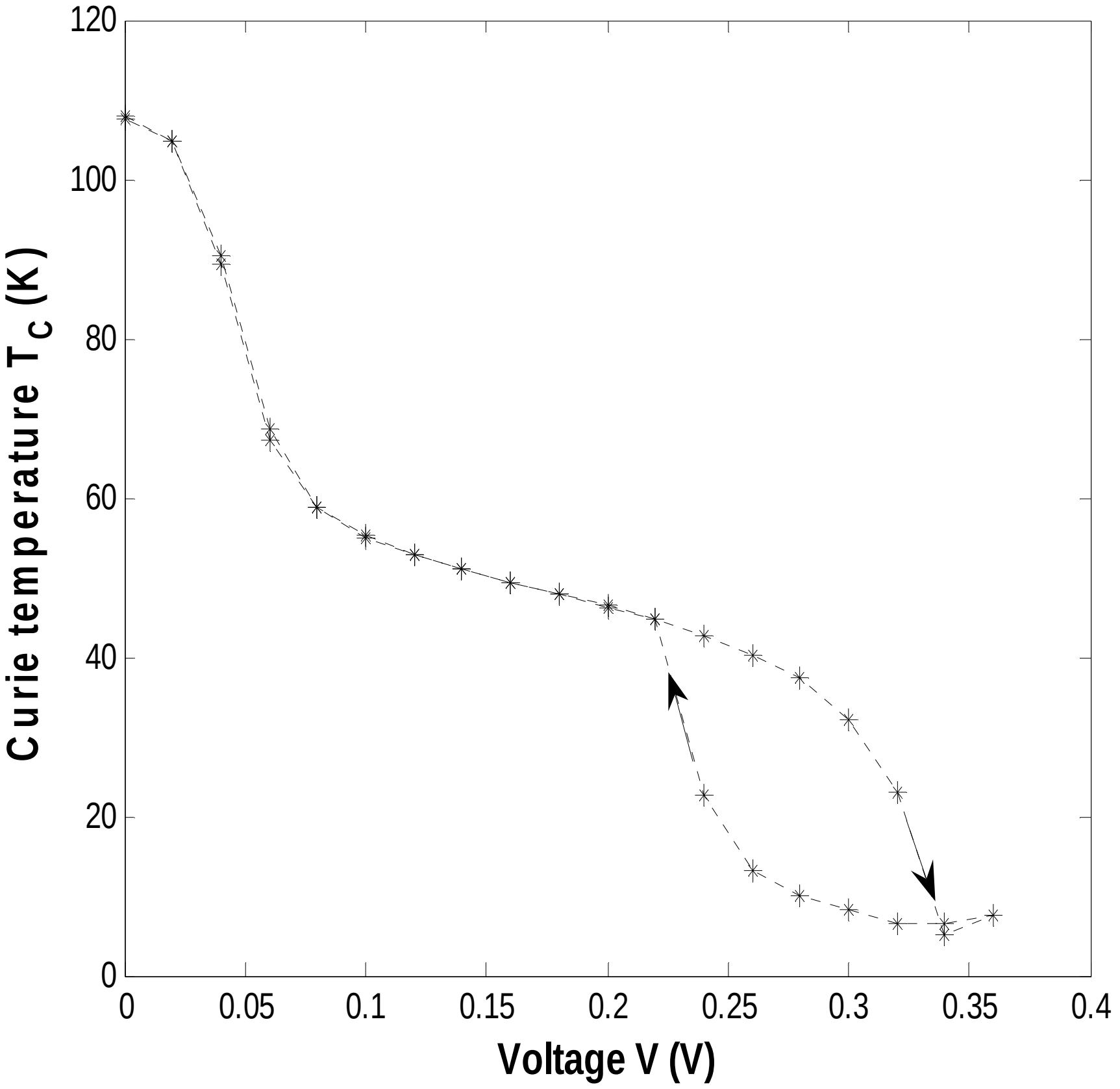

FIG. 3. Hysteretic Curie temperature vs. voltage ($T_C$-V) characteristics of the ferromagnetic well resonant tunneling diode.